%% file: main.tex
\documentclass[aps,prl,reprint,superscriptaddress]{revtex4-2}

\usepackage{silence}
\usepackage{graphicx}
\usepackage{amsmath}
\usepackage{amsthm}
\usepackage{amssymb}
\usepackage{dcolumn}
\usepackage{bm}
\usepackage{multirow}
\usepackage{empheq}
\usepackage{tikz}
\usepackage{enumitem}
\usepackage{booktabs}
\usepackage{xspace}
\usepackage[colorlinks=true, allcolors=blue]{hyperref}
\usepackage[nameinlink]{cleveref}
\usepackage{bbold}
\usepackage{array}
\usepackage{microtype}

\newcommand{\chsh}[0]{2.8017}
\input{macros}

\begin{document}


\title {Dynamical Readout of Measurement Statistics and Emergent Entanglement-Like States in Classical Networks}

\author{Songbo Xie}
    \affiliation{Department of Electrical and Computer Engineering, North Carolina State University, Raleigh, North Carolina 27606, USA}
\author{Ethan Dickey}
    \affiliation{Department of Computer Science, Purdue University, West Lafayette, Indiana 47907, USA}
\author{Syed Ahmed Taimoor}
    \affiliation{Department of Computer Science, North Carolina State University, Raleigh, North Carolina 27606, USA}
\author{Sabre Kais}
\email{skais@ncsu.edu}
    \affiliation{Department of Electrical and Computer Engineering, North Carolina State University, Raleigh, North Carolina 27606, USA}
\affiliation{ Department of Chemistry and Department of Physics, North Carolina State University, Raleigh, North Carolina 27606, USA}

\begin{abstract}

Can a classical network not only encode a quantum-like state, but also read out its measurement statistics through its own collective dynamics? Here we introduce a network-native readout scheme for a structured classical network whose community modes form an effective two-qubit state space. Network connectivity selects a dominant collective mode that encodes the state, while a fixed set of ten elementary connectivity perturbations probes its collective response. The resulting shifts of the dominant growth rate form a complete basis of expectation values for the real two-qubit sector, from which joint-outcome probabilities associated with arbitrary real projectors are reconstructed by linear combination. Once these ten spectral responses are measured, the same data generate correlations over a continuous family of measurement settings. As a benchmark, for an encoded Bell state four selected correlations give $|S|=\chsh>2$, whereas a separable product-state reference remains within the Clauser--Horne--Shimony--Holt (CHSH)  bound $|S| \leq2 $. These results establish an informationally complete dynamical readout of effective two-qubit states directly from the spectral response of a classical network.
\end{abstract}

\maketitle


{\it Introduction}.---Quantum information is usually formulated in terms of microscopic quantum degrees of freedom (DOF), but closely related structures can also arise in classical systems. In classical optics, distinct DOF can support nonseparable states and even Bell-type correlations \cite{spreeuw1998classical,spreeuw2001classical,qian2011entanglement,kagalwala2013bell,aiello2015quantum,qian2015shifting,borges2010bell,mclaren2015measuring,simon2010nonquantum,ndagano2017characterizing}. Acoustic and elastic-wave systems provide another realization, where coupled modes form nonseparable and qubit-like state spaces \cite{deymier2017non,hasan2019sound,hasan2021experimental}. More recently, collective modes of synchronizing networks have been developed into quantum-like state spaces supporting effective bits, tensor-product structures, gate operations, interference, and entangled-state dynamics \cite{Scholes2024QuantumLike,AmatiScholes2025Encoding,AmatiScholes2025,scholes2025product,Scholes2026Dynamics}. Within this network setting, however, how joint measurement statistics can be extracted directly from the underlying collective dynamics remains an open question.

Network connectivity provides a natural route to such a readout. Symmetry and balanced connectivity can make the uniform modes of different communities form a low-dimensional invariant subspace of the network dynamics \cite{pecora2014cluster,schaub2016graph,siddique2018symmetry,o2013observability}. Each basis state is represented by the uniform collective mode of an entire community rather than by an individual node. Under linear dynamics, the mode associated with the largest eigenvalue is naturally selected at long times, defining a collective quantum-like state whose form can be engineered through the network connectivity \cite{DickeyKais2026Complex,DickeyVyasKais2025}. This raises a more operational question: can the same classical network that encodes such a state also read out its measurement statistics through its own collective response?

To probe this collective state, we use the network connectivity itself as a measurement interface. We introduce ten elementary connectivity perturbations. For each perturbation $M_\mu$, the Hellmann--Feynman relation gives the first-order spectral response $\delta\lambda_\mu\simeq\langle\psi_G|M_\mu|\psi_G\rangle$ \cite{feynman1939forces}. These ten responses form a complete basis for reconstructing any real symmetric joint-outcome projector on the effective two-qubit space, so that the probabilities $p(a,b|x,y)$ are obtained as linear combinations of the measured $\delta\lambda_\mu$. Each shift is read out dynamically from the change in the asymptotic growth rate of the network state, without diagonalizing the adjacency matrix explicitly. Once these elementary responses are known, the same data reconstruct joint probabilities and correlations over a continuous family of measurement settings. We use the CHSH functional as a benchmark of this readout: for an encoded Bell state, four reconstructed correlations give $|S|=\chsh>2$, while a separable product-state reference remains within the CHSH bound.

We emphasize that the CHSH construction is used here as a benchmark of the reconstructed measurement statistics, not as a test of Bell nonlocality \cite{bell1964einstein,brunner2014bell,korolkova2024operational,pereira2014quantum,Kais-Enyanglemnet}: the effective Alice--Bob measurements are not implemented as independent local measurements on spatially separated subsystems, nor does the network realize measurement-induced state reduction. In this statistical sense, our framework extends the quantum--classical correspondence beyond state representation to measurement statistics encoded directly in the collective dynamics of a classical network.

\begin{figure}
    \centering
    \includegraphics[height=0.45\linewidth]{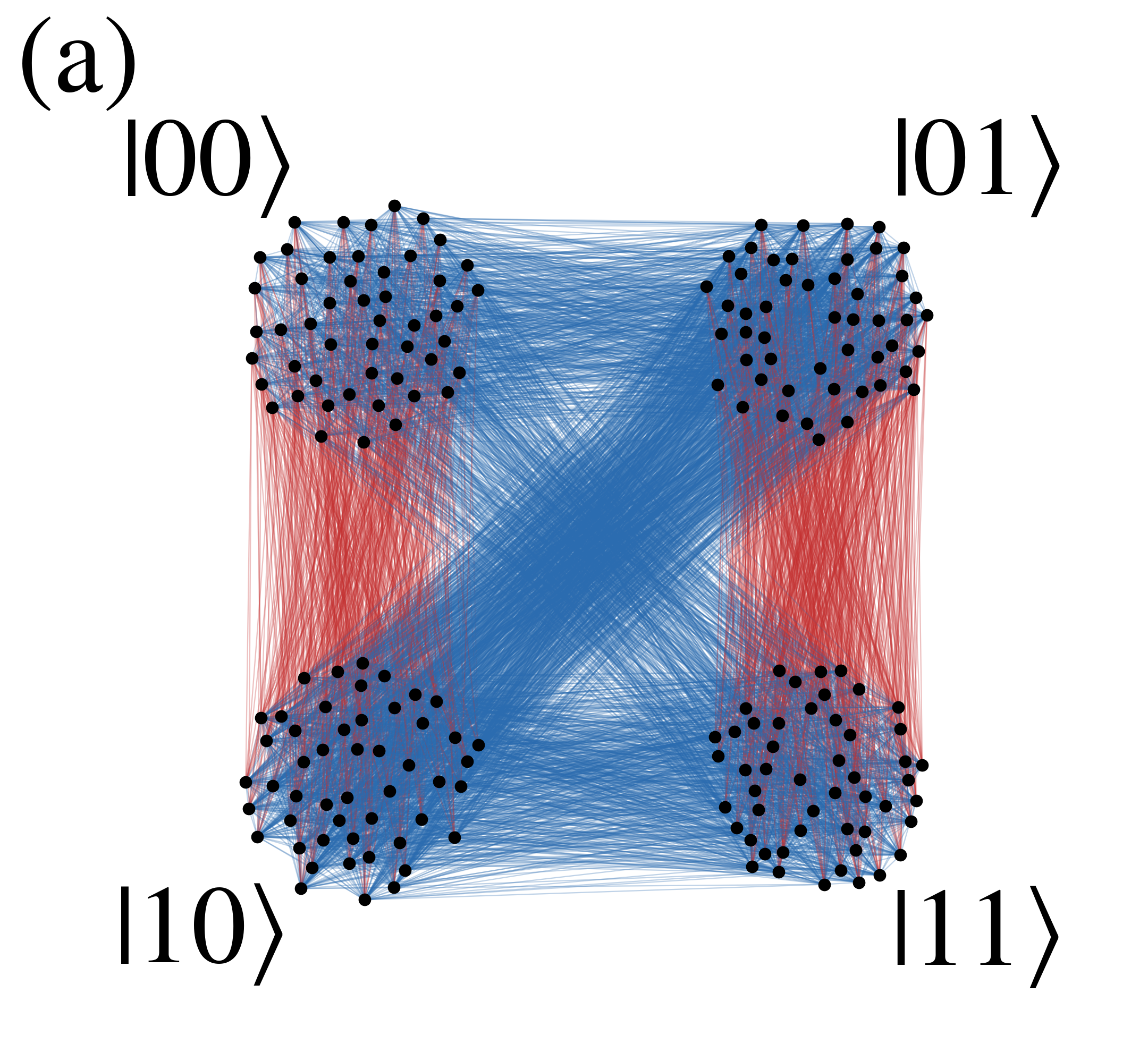}
    \includegraphics[height=0.45\linewidth]{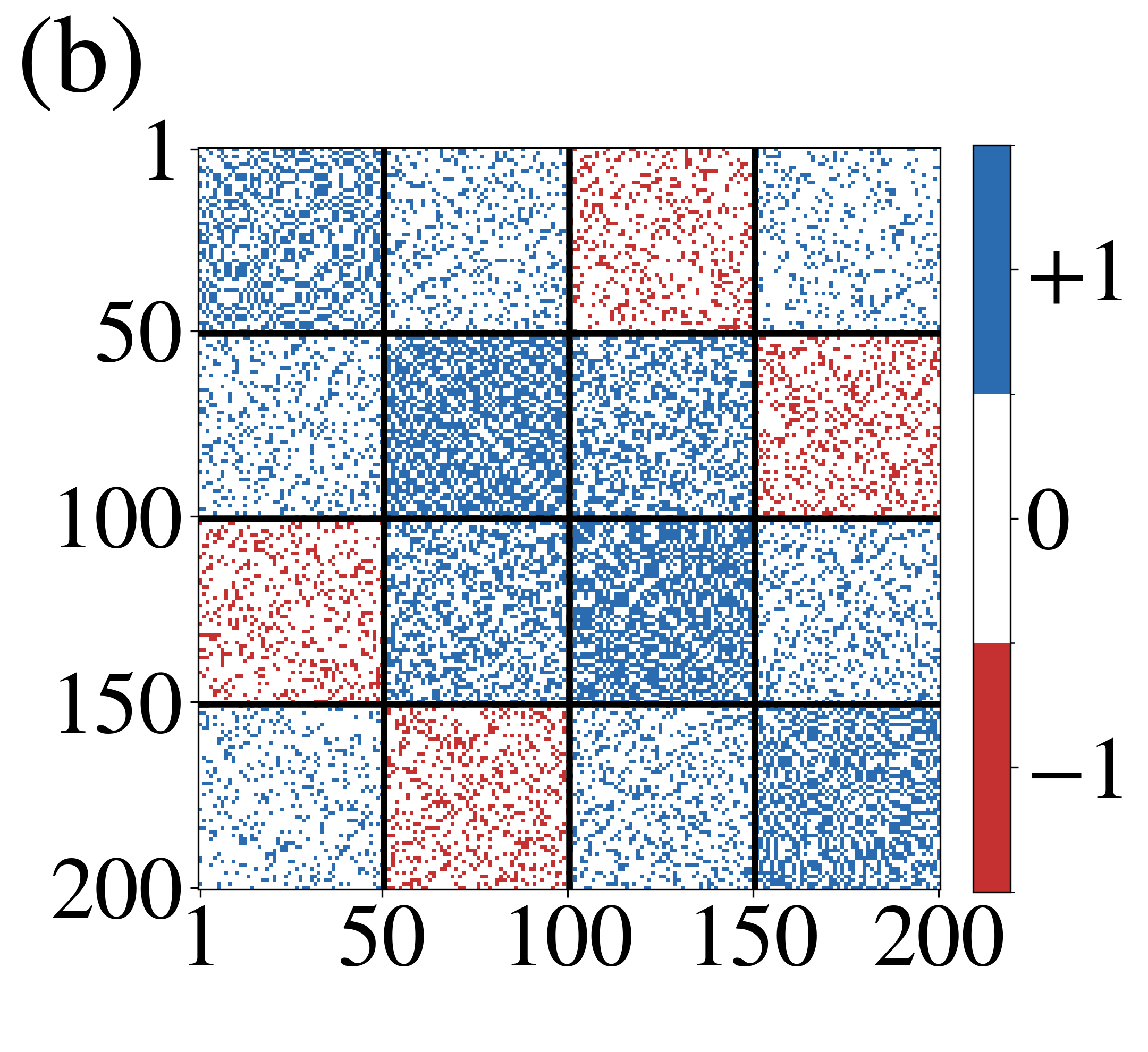}
    \includegraphics[height=0.36\linewidth]{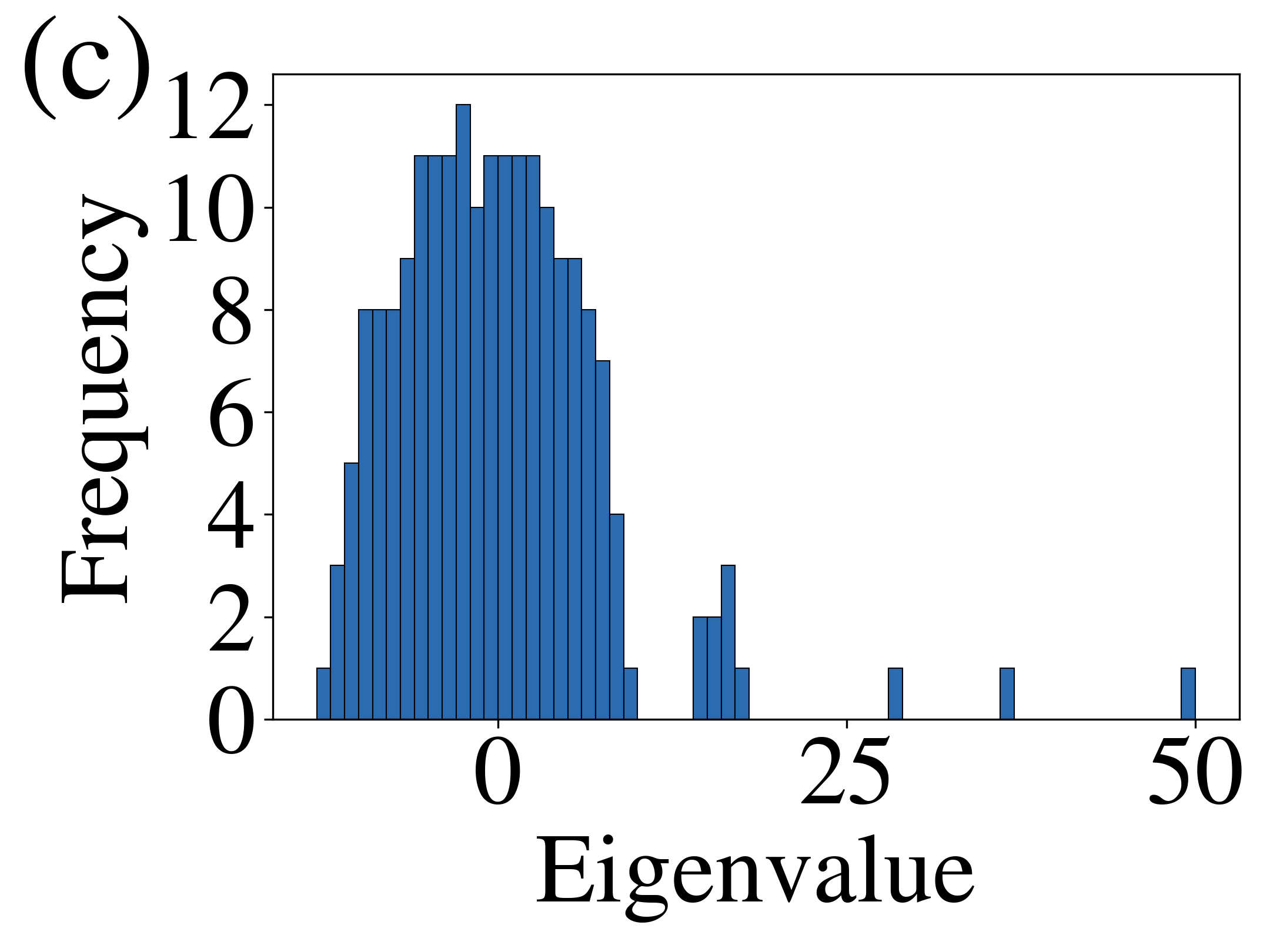}
    \includegraphics[height=0.36\linewidth]{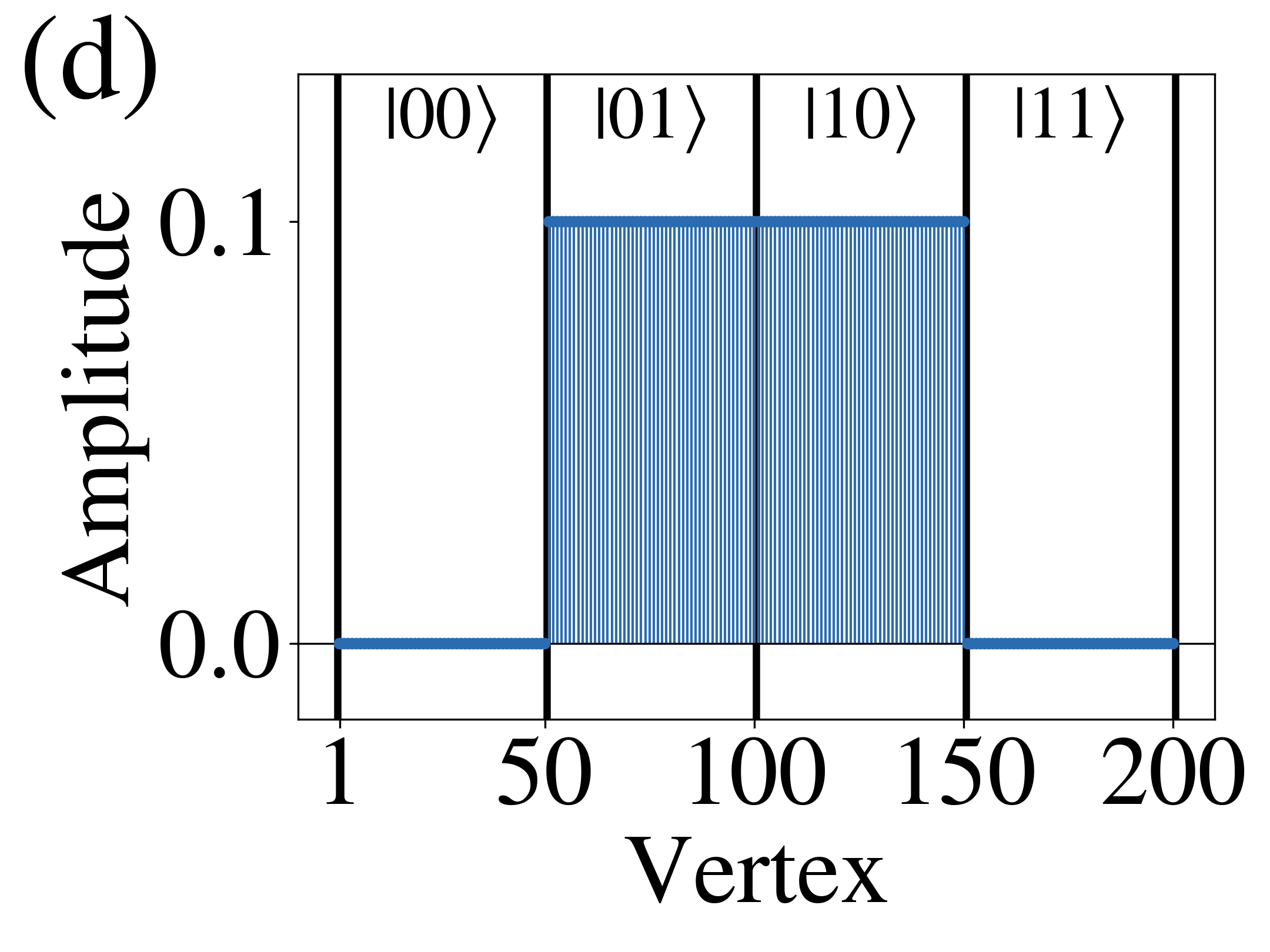}
    \caption{\textbf{Bell-like collective state in a classical network.}
        The communities $(\kzz,\kzo,\koz,\koo)$ each contain $n=50$ vertices, with internal regularities $(k_1,k_2,k_3,k_4)=(20,30,30,25)$ and signed inter-community biregularities $(l_{12},l_{13},l_{14},l_{23},l_{24},l_{34})=(10,-10,8,20,-12,12)$.
        \textbf{(a)} Graph $G$, with positive (blue) and negative (red) edges.
        \textbf{(b)} Adjacency matrix $A(G)$, with black community boundaries and blue, red, and white entries denoting $+1$, $-1$, and $0$.
        \textbf{(c)} Spectrum, with isolated, nondegenerate dominant eigenvalue $\lambda_{\max}=50$.
        \textbf{(d)} Normalized dominant eigenvector: amplitudes vanish on $\kzz$ and $\koo$ and are equal on $\kzo$ and $\koz$, identifying $\kbellp=(\kzo+\koz)/\sqrt{2}$.}
        
    \label{fig:state}
\end{figure}

{\it Collective states in a classical network.---}We consider a graph $G$ with four interacting communities $V_i$, each containing $n$ vertices and defining a normalized uniform collective mode
\begin{equation}
    \mathbf u_i=\frac{1}{\sqrt{n}}\sum_{a\in V_i}\mathbf e_a, \qquad i=1,2,3,4,
\end{equation}
where $\mathbf e_a$ is a unit amplitude on vertex $a$. Each $\mathbf u_i$ vector is uniform on $V_i$ and zero elsewhere. These orthonormal modes span the code subspace $\mathcal C_G=\operatorname{span}\{\mathbf u_1,\mathbf u_2,\mathbf u_3,\mathbf u_4\}$.

Identifying $(\mathbf u_1,\mathbf u_2,\mathbf u_3,\mathbf u_4)$ with $(|00\rangle,|01\rangle,|10\rangle,|11\rangle)$ gives $\mathcal C_G$ an effective two-qubit structure. For the real-symmetric networks considered here, the collective states occupy the real sector of $\mathbb C_A^2\otimes\mathbb C_B^2$, with the first and second labels defining Alice and Bob, respectively.

The edges determine which state in this space is selected. The real symmetric adjacency matrix $A(G)$ has entries $A_{ab}=+1,-1,0$ for positive, negative, or absent edges, respectively. Within each community $i$, all edges are positive, and every vertex has $k_i$ neighbors in that community. Between communities $i$ and $j$, every vertex has $|l_{ij}|$ neighbors in the other community. All edges within the corresponding off-diagonal block have the same sign, determined by the sign of $l_{ij}$.


For the graph family used here, these connectivity conditions are chosen such that the largest eigenvalue $\lambda_{\max}$ of $A(G)$ is nondegenerate and separated from the rest of the spectrum. The corresponding top eigenvector lies in $\mathcal C_G$ and has a general form $\boldsymbol{\psi}_{\max}=\alpha\mathbf u_1+\beta\mathbf u_2+\gamma\mathbf u_3+\delta\mathbf u_4$ (see Sec.~I in \cite{supplemental}).

Rather than diagonalizing $A(G)$ and selecting its top eigenvector explicitly, we let this collective mode emerge from the network dynamics. We take $H_G=-A(G)$ in a Schr\"odinger-like equation $i\frac{d}{dt}\mathbf v(t)=H_G\mathbf v(t)$. Under the imaginary-time transformation $t=-i\tau$, it becomes
\begin{equation}\label{dynamics}
    \frac{d}{d\tau}\mathbf v(\tau)=A(G)\mathbf v(\tau).
\end{equation}
This evolution converts the eigenvalues of $A(G)$ into exponential growth rates, so that the spectrally isolated dominant mode is dynamically selected at long times.

An arbitrary initial mode can be decomposed as $\mathbf v(0)=\sum_m c_m\boldsymbol{\psi}_m$, where $A(G)\boldsymbol{\psi}_m=\lambda_m\boldsymbol{\psi}_m$, giving $\mathbf v(\tau)=\sum_m c_m e^{\lambda_m\tau}\boldsymbol{\psi}_m$. If $c_{\max}\neq0$, the spectral gap suppresses all other modes. After normalization,
\begin{equation}
    \lim_{\tau\rightarrow\infty}\frac{\mathbf v(\tau)}{\|\mathbf v(\tau)\|} = \boldsymbol{\psi}_{\max}.
\end{equation}
Thus, the long-time dynamics select a unique collective mode, identified with the encoded two-qubit state
\begin{equation}
    |\psi_G\rangle =
    \alpha\kzz+
    \beta\kzo+
    \gamma\koz+
    \delta\koo,
\end{equation}
with amplitudes determined by the intra-community connectivities $k_i$ and inter-community connectivities $l_{ij}$.

As an example, Fig.~\ref{fig:state} shows a four-community network together with its adjacency matrix, eigenvalue spectrum, and dominant eigenvector. The corresponding encoded state is the Bell state
\begin{equation}\label{bellstate}
    \kbellp=\frac{1}{\sqrt{2}}\Big(\kzo+\koz\Big).
\end{equation}


\begin{figure*}
    \centering
    \raisebox{0.04\linewidth}{\includegraphics[height=0.22\linewidth]{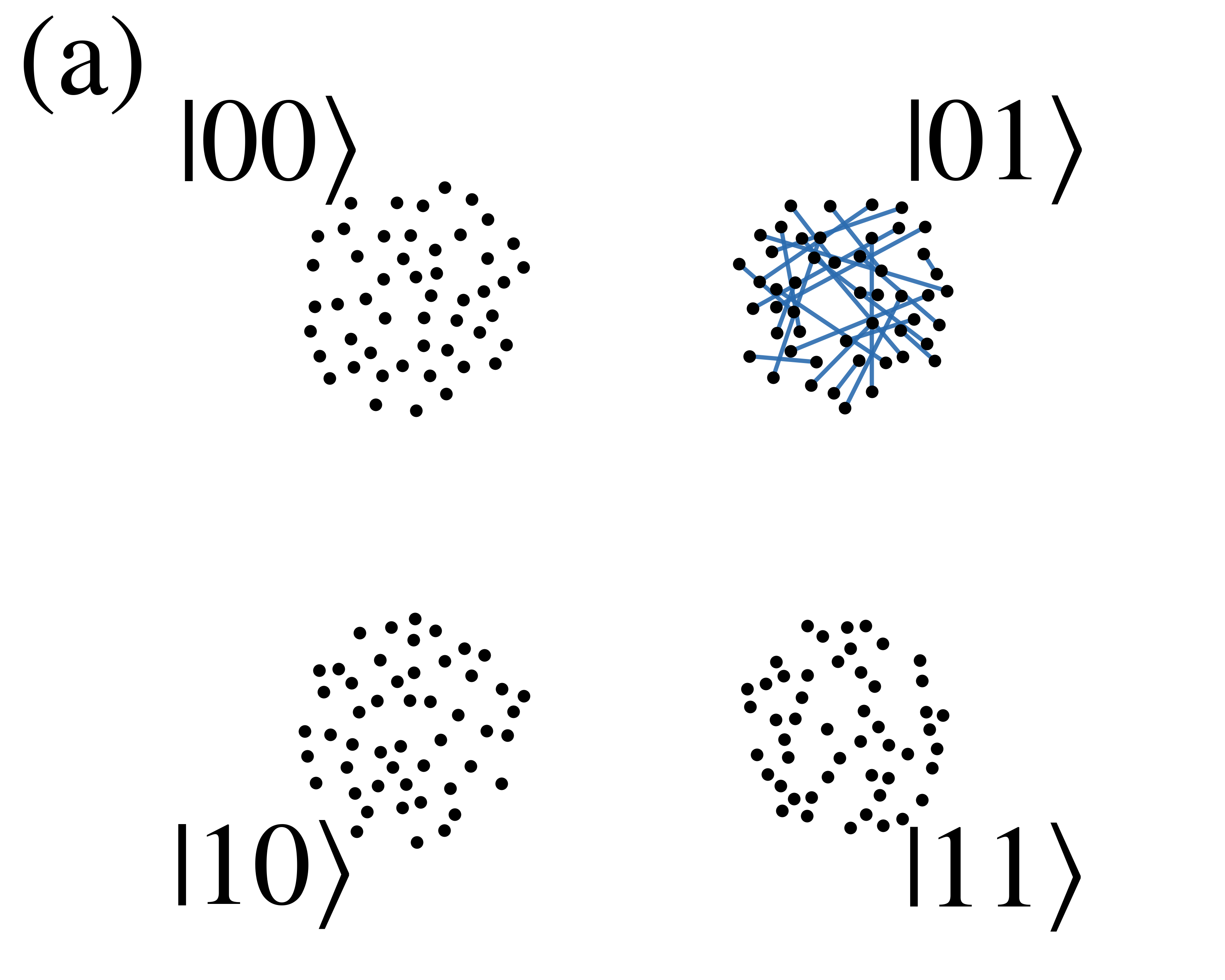}}
    \includegraphics[height=0.26\linewidth]{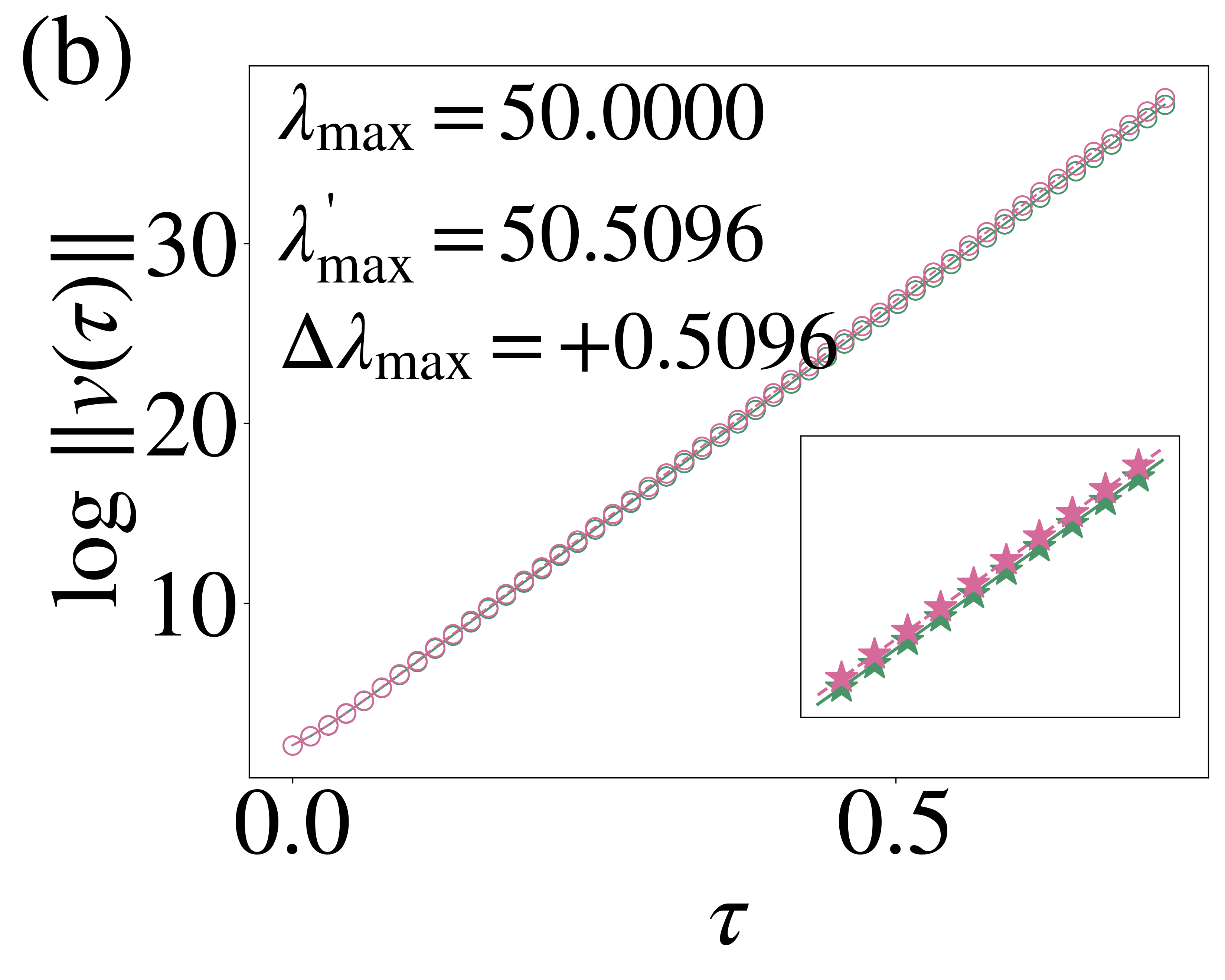}\\    
    \raisebox{-0.5\height}{\includegraphics[height=0.2\linewidth]{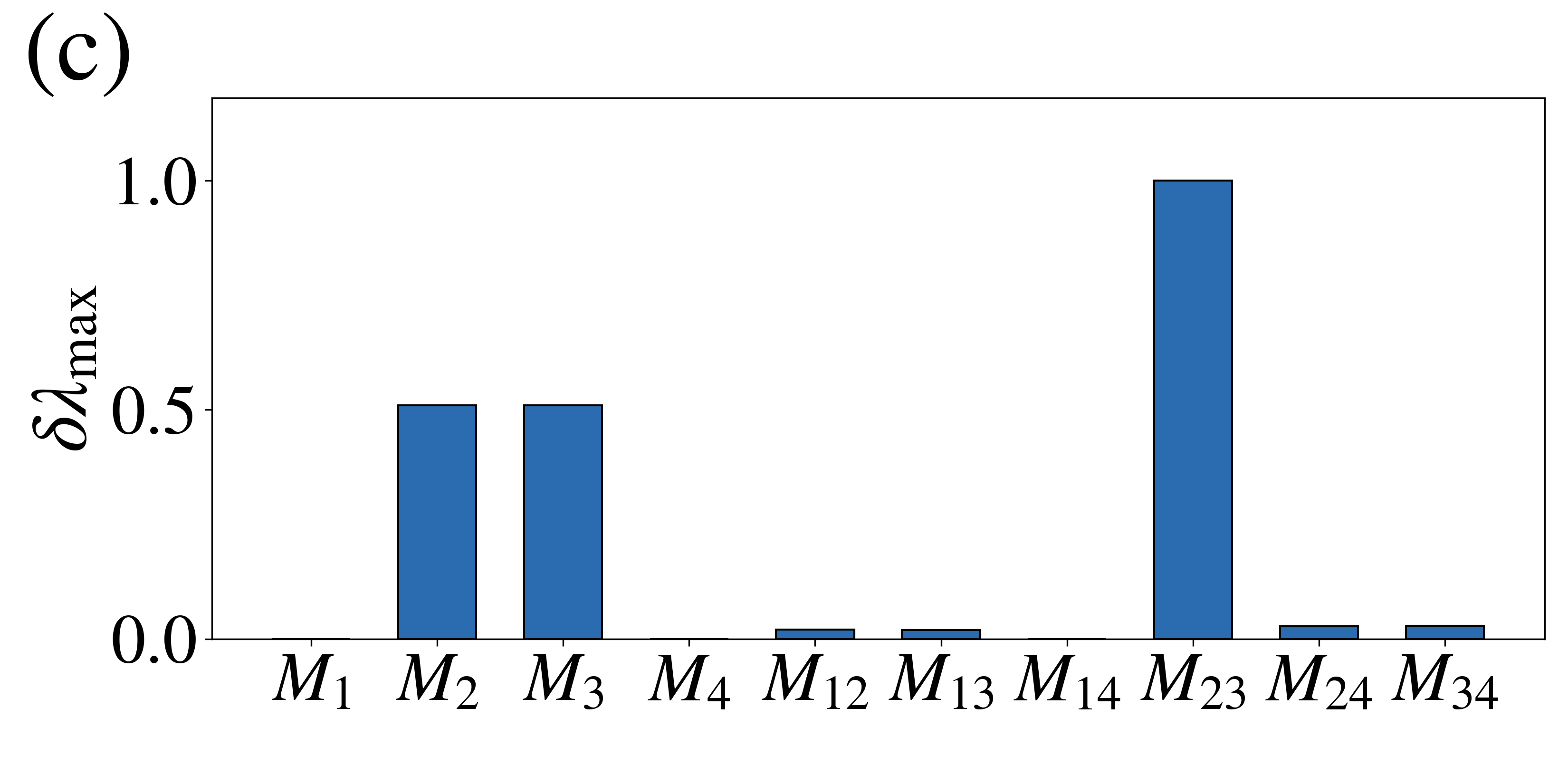}} \raisebox{-0.5\height}{\includegraphics[height=0.2\linewidth]{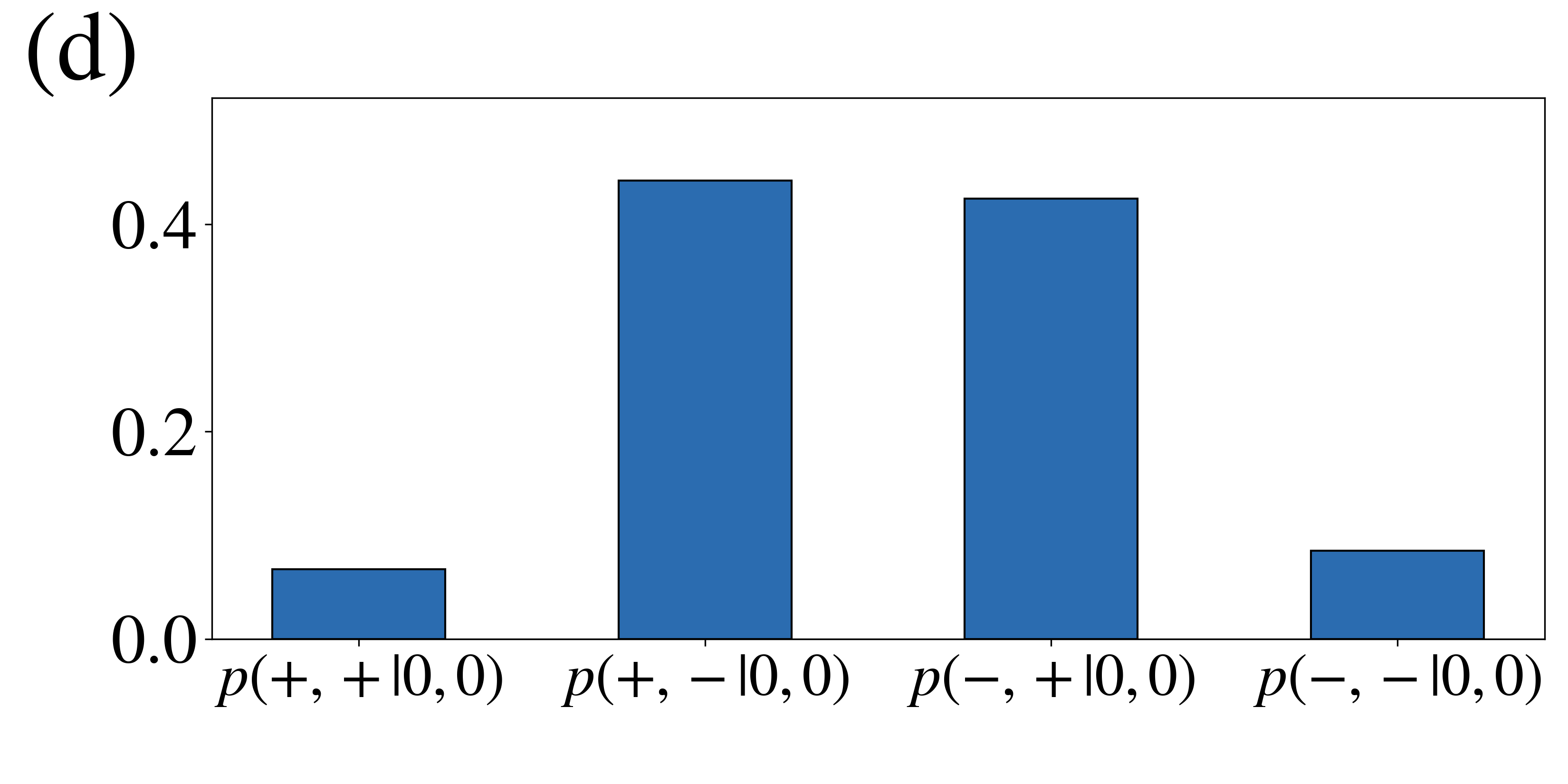}}
    \caption{\textbf{Joint-probability readout from connectivity perturbations.}
    \textbf{(a)} Perturbation $M_2$ increases only the internal regularity of $\kzo$ by one.
    \textbf{(b)} $\log\lVert\mathbf v(\tau)\rVert$ for unperturbed $\dot{\mathbf v}=A\mathbf v$ (green) and perturbed dynamics (pink), starting from the same random state. The late-time slope changes from $\lambda_{\max}=50.0000$ to $\lambda'_{\max}=50.5096$, giving $\delta\lambda_2=0.5096$.
    \textbf{(c)} Growth-rate shifts for all ten perturbations. Only $M_2$, $M_3$, and $M_{23}$ respond, reflecting $\kbellp$.
    \textbf{(d)} Joint probabilities $p(a,b|0,0)$ for $A_0=Z$ and $B_0=\onesqrttwo(Z-X)$, reconstructed from (c) using the projector coefficients. Their signed sum gives $\braketnom{A_0}{B_0}=-0.7011$, close to the ideal $-\onesqrttwo\approx-0.7071$.}
    
    \label{fig:readout}
\end{figure*}

{\it Joint-outcome readout.---}Alice and Bob choose dichotomic observables $A_x$ and $B_y$, respectively, with outcomes $a,b=\pm1$ and projectors
\begin{equation}
    \Pi^{A}_{a|x}=\frac{I+aA_x}{2}, \qquad
    \Pi^{B}_{b|y}=\frac{I+bB_y}{2},
\end{equation}
with the joint-outcome projector $\Pi^{AB}_{ab|xy} = \Pi^{A}_{a|x}\otimes\Pi^{B}_{b|y}.$ For the collective state $|\psi_G\rangle$, the probability of obtaining the joint outcome $(a,b)$ is
\begin{equation}\label{jointprobability}
    \widetilde p(a,b|x,y) = \langle\psi_G|\Pi^{AB}_{ab|xy}|\psi_G\rangle.
\end{equation}

However, a generic joint-outcome projector cannot be implemented as a single connectivity perturbation. For the measurement family considered here, $\Pi^{AB}_{ab|xy}$ is a real symmetric $4\times4$ matrix whose entries generally contain irrational ratios. Instead, we reconstruct it from a fixed set of elementary connectivity perturbations. Since the space of real symmetric $4\times4$ matrices is ten dimensional, ten independent perturbations are sufficient.

We choose ten basis operators on the code subspace $\mathcal{C}_G$: four diagonal operators $M_i=|i\rangle\langle i|$, with $i=1,2,3,4$, and six symmetric off-diagonal operators $M_{ij}=|i\rangle\langle j|+|j\rangle\langle i|$, with $i<j$. These operators form a complete basis for the real symmetric $4\times4$ matrices. Their expectation values provide an informationally complete tomographic readout of the real two-qubit sector \cite{james2001measurement,toninelli2019concepts}. Accordingly, any real joint-outcome projector can be expanded as (see Sec.~II and Sec.~III of \cite{supplemental})
\begin{equation}
    \Pi^{AB}_{ab|xy}
    =
    \sum_{\mu=1}^{10}c^{(\mu)}_{ab|xy}M_\mu,
\end{equation}
where the coefficients $c^{(\mu)}_{ab|xy}$ are fixed by the chosen measurement settings.

Each $M_\mu$ is implemented by an adjacency-matrix perturbation $\delta A_\mu$ on the full network. For $M_i$, only the internal regularity of community $i$ is changed, with $\delta\, k_i=1$; for $M_{ij}$, only the signed biregularity between communities $i$ and $j$ is changed, with $\delta\, l_{ij}=1$. All other connectivities remain unchanged. This gives $\left.\delta A_\mu\right|_{\mathcal{C}_G}=M_\mu$.

Provided that the perturbation is small relative to the spectral separation of the dominant mode, first-order eigenvalue perturbation theory gives \cite{feynman1939forces,stewart1990matrix} (see also Sec.~IV of \cite{supplemental})
\begin{equation}
    \begin{split}
        \delta\lambda_\mu
        \equiv {}&
        \lambda_{\max}\!\left[A(G)+\delta A_\mu\right]
        -\lambda_{\max}\!\left[A(G)\right]\\
        \simeq {}&
        \boldsymbol{\psi}_{\max}^{T}\delta A_\mu\boldsymbol{\psi}_{\max}
        =
        \langle\psi_G|M_\mu|\psi_G\rangle.
    \end{split}
\end{equation}

Importantly, the shifts $\delta\lambda_\mu$ can be obtained directly from the network dynamics, without diagonalizing the adjacency matrix. Under Eq.~\eqref{dynamics}, one obtains
\begin{equation}
    \begin{split}
        \delta\lambda_\mu = \lim_{\tau\rightarrow\infty} \frac{1}{\tau}
    \log\left(\dfrac{\left\|e^{[A(G)+\delta A_\mu]\tau}\mathbf v(0)\right\|}{\left\|e^{A(G)\tau}\mathbf v(0)\right\|}\right), &  \\
    \end{split}
\end{equation}
for any initial state with nonzero overlap with the dominant mode. The ten measured shifts therefore provide the expectation values of the ten basis operators.

Starting from Eq.~\eqref{jointprobability}, for a given setting pair $(x,y)$ and outcome $(a,b)$, their linear combination reconstructs the corresponding joint-outcome probability,
\begin{equation}
    \begin{split}
        \widetilde p(a,b|x,y) = \langle\psi_G|\Pi^{AB}_{ab|xy}|\psi_G\rangle =
        \sum_{\mu=1}^{10} c^{(\mu)}_{ab|xy}\,\delta\lambda_\mu.
    \end{split}
\end{equation}
Finite perturbations introduce higher-order corrections, causing the four reconstructed values to deviate slightly from unit normalization. We therefore use
\begin{equation}
    p(a,b|x,y)
    =
    \frac{\widetilde p(a,b|x,y)}
    {\displaystyle\sum_{a',b'=\pm1}\widetilde p(a',b'|x,y)}.
\end{equation}
Thus, once the ten elementary spectral shifts are measured, joint probabilities for any measurement setting are obtained by changing the coefficients $c^{(\mu)}_{ab|xy}$.

As a representative example, for $\kbellp$ with $A_0=Z$ and $B_0=\onesqrttwo(Z-X)$, the projector coefficients combine the ten shifts into $p(a,b|0,0)$, as illustrated in Fig.~\ref{fig:readout}.



\begin{figure}
    \centering
    \includegraphics[width=0.85\linewidth]{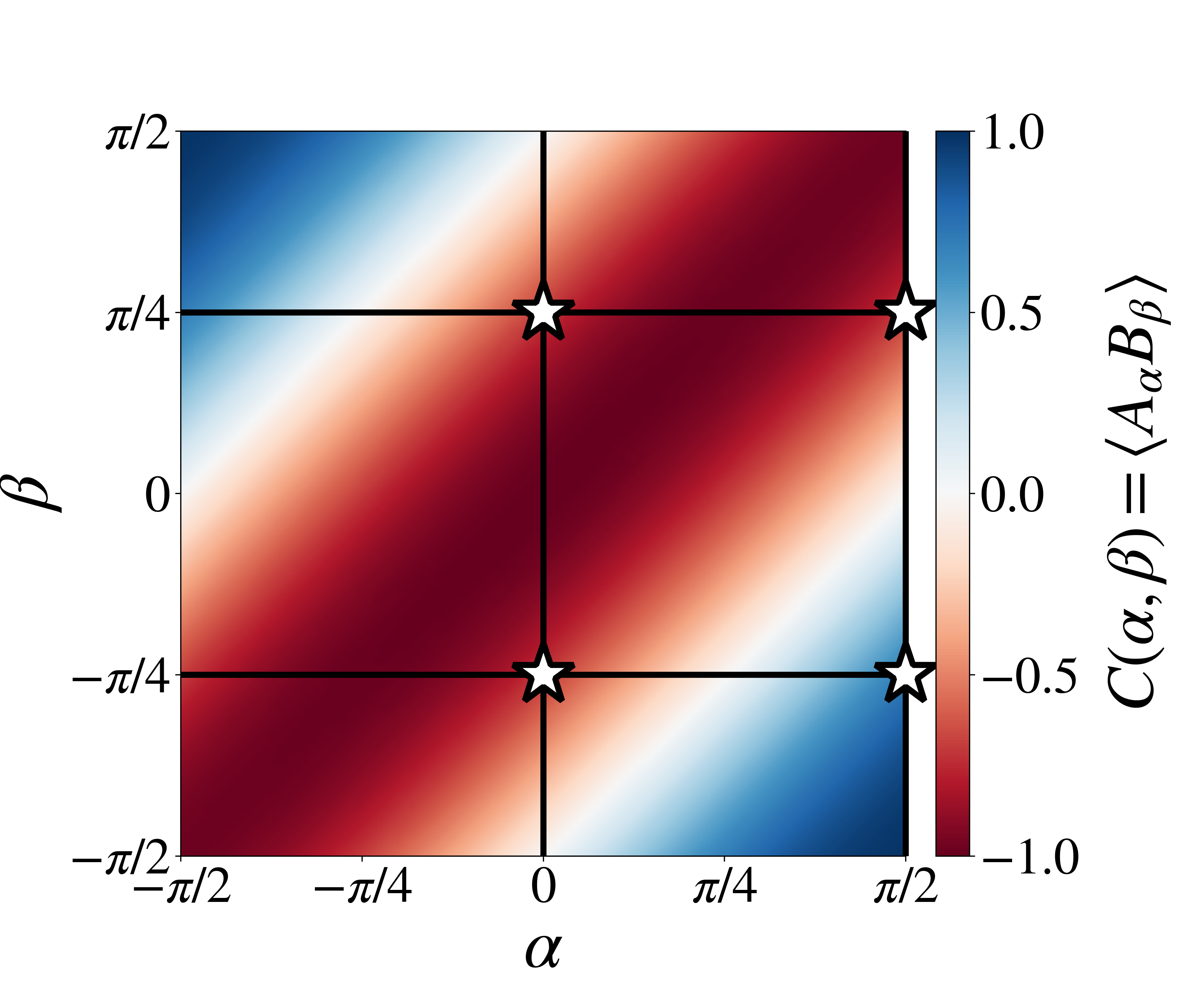}
    \caption{\textbf{Correlation landscape of the encoded Bell state.}
    The correlation $C(\alpha,\beta)=\langle A_\alpha B_\beta\rangle$ reconstructed over the full range of measurement settings, with $A_\alpha=\cos(\alpha)\,Z+\sin(\alpha)\,X$ and $B_\beta=\cos(\beta)\,Z-\sin(\beta)\,X$. Each point is obtained by combining the same ten elementary growth-rate shifts $\delta\lambda_\mu$ measured in Fig.~2(c) with the coefficients of the corresponding joint-outcome projectors. Blue and red denote positive and negative correlations, respectively. The resulting bands follow $\alpha-\beta=\mathrm{const}$, reproducing the ideal dependence $-\cos(\alpha-\beta)$ for $\kbellp$. Black lines mark the CHSH settings $\alpha_0=0$, $\alpha_1=\pi/2$, $\beta_0=\pi/4$, and $\beta_1=-\pi/4$; the stars indicate the four correlations entering $S$, yielding $|S|=\chsh>2$.}
    \label{fig:heat}
\end{figure}

{\it Correlation reconstruction and CHSH benchmark.---}With the collective state and elementary spectral responses established, we now reconstruct Bell correlations over a continuous family of measurement settings. We parameterize Alice's and Bob's observables in the $X$-$Z$ plane as
\begin{equation}
    \begin{split}
        A(\alpha)=&\cos(\alpha)\,Z+\sin(\alpha)\,X,\\
        B(\beta)=&\cos(\beta)\,Z-\sin(\beta)\,X.
    \end{split}
\end{equation}
For each setting pair $(\alpha,\beta)$, the four joint-outcome projectors determine the coefficients $c^{(\mu)}_{ab|\alpha\beta}$ that combine the same ten measured shifts $\delta\lambda_\mu$ into the probabilities $p(a,b|\alpha,\beta)$. The corresponding correlation is then
\begin{equation}
C(\alpha,\beta)
=
\sum_{a,b=\pm1}ab\,p(a,b|\alpha,\beta).
\end{equation}
Varying $(\alpha,\beta)$ reconstructs the full correlation landscape.

\begin{figure}
    \centering
    \includegraphics[height=0.45\linewidth]{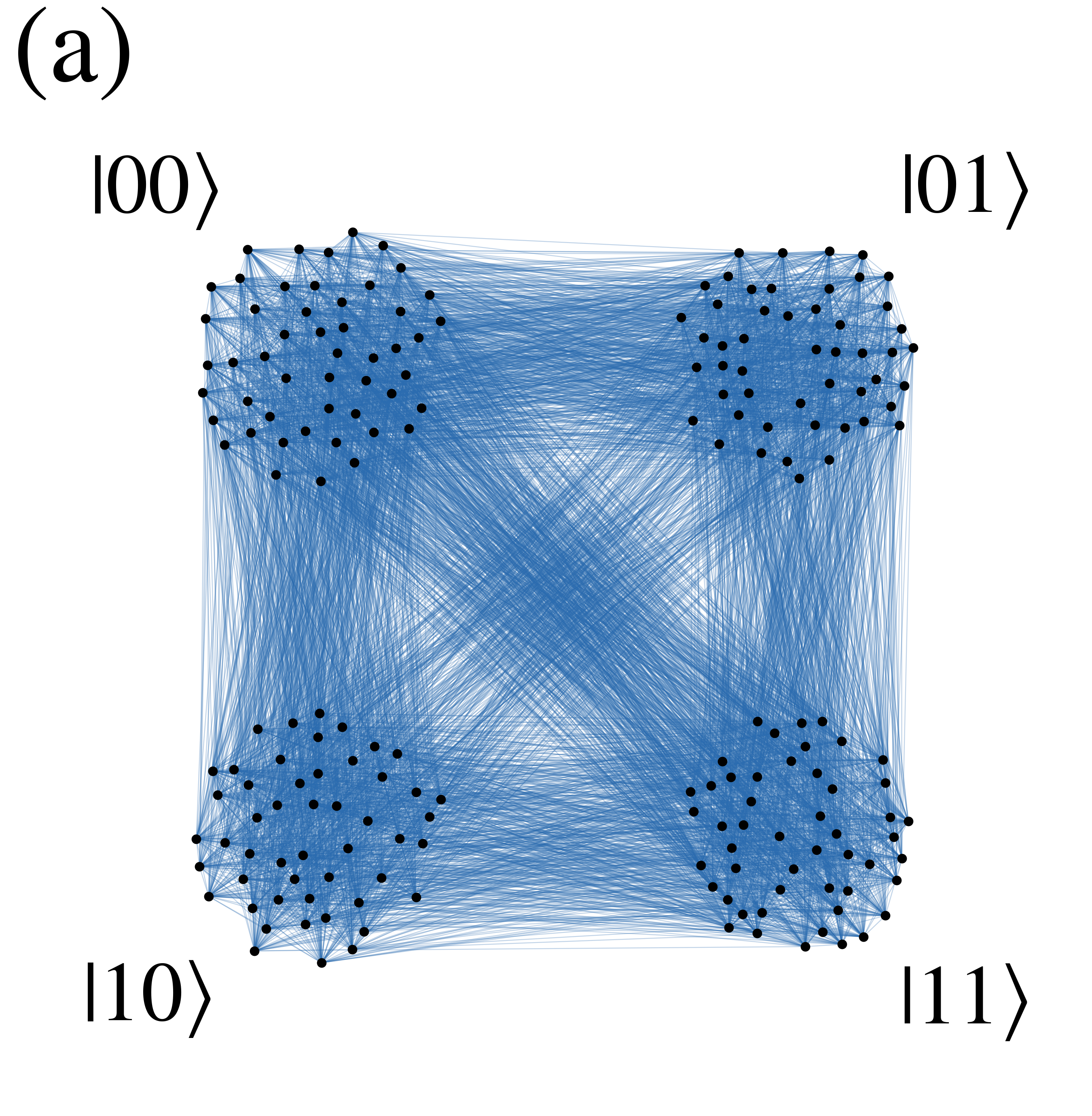}
    \includegraphics[height=0.45\linewidth]{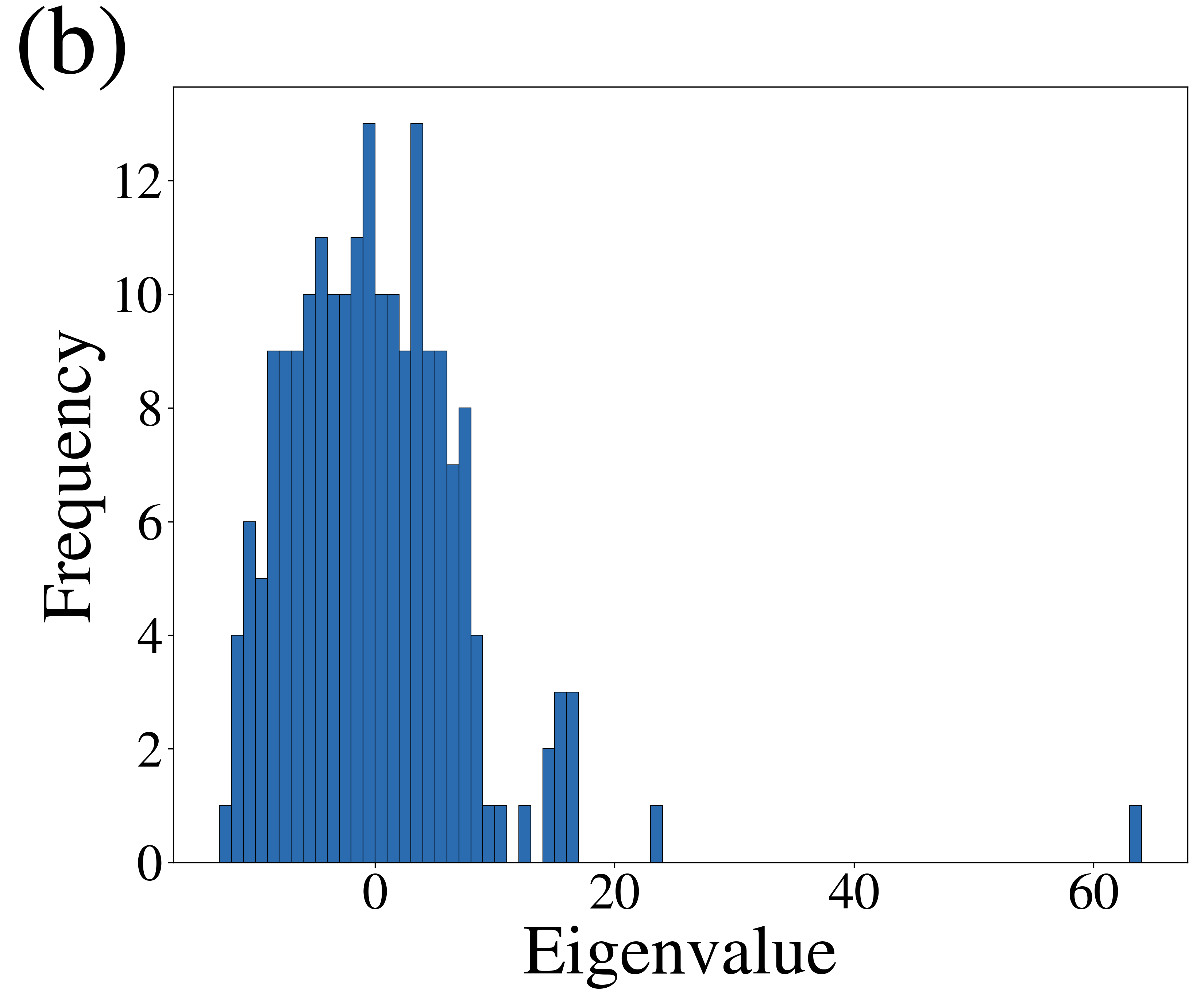}
    \includegraphics[height=0.45\linewidth]{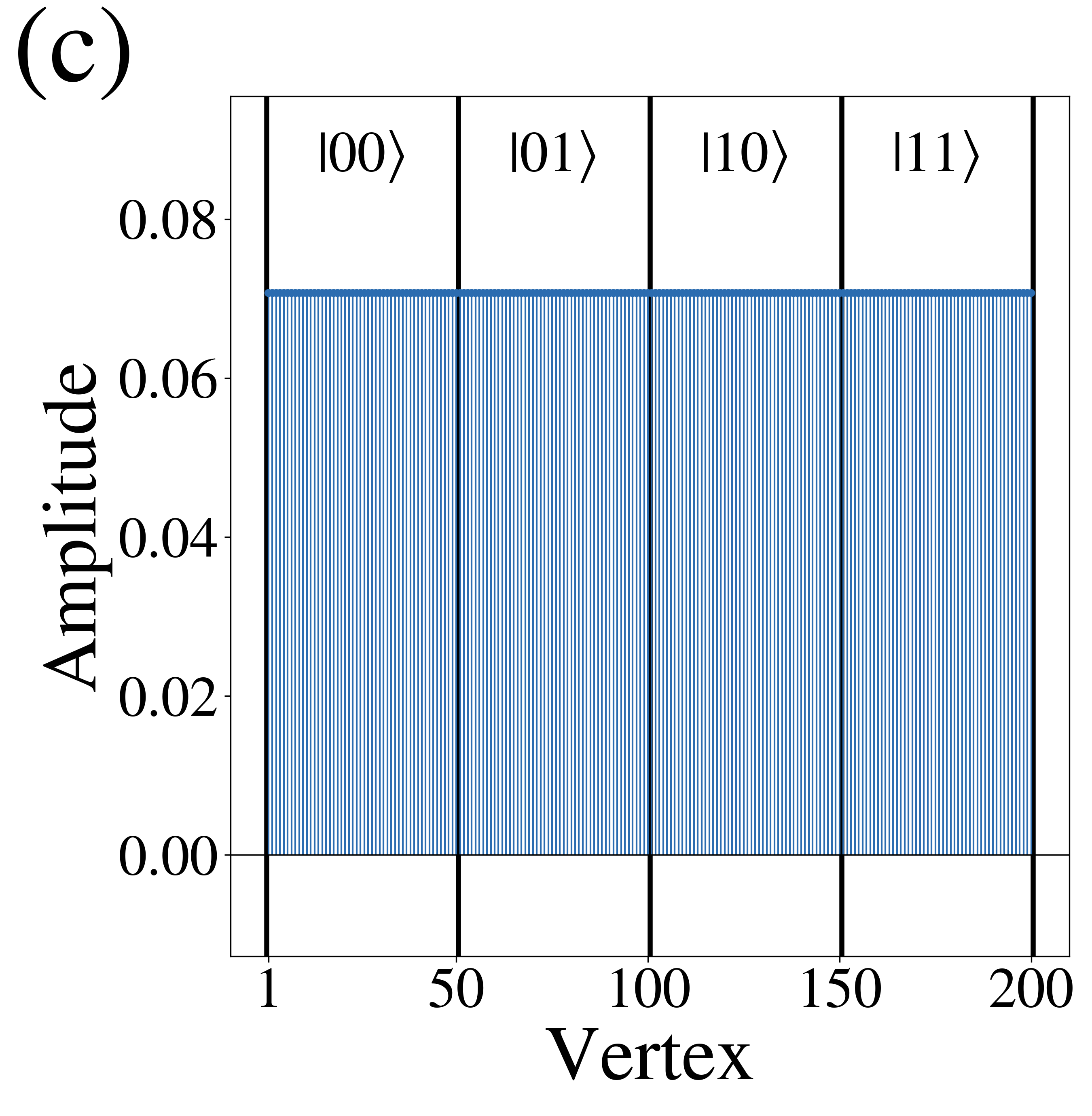}
    \includegraphics[height=0.45\linewidth]{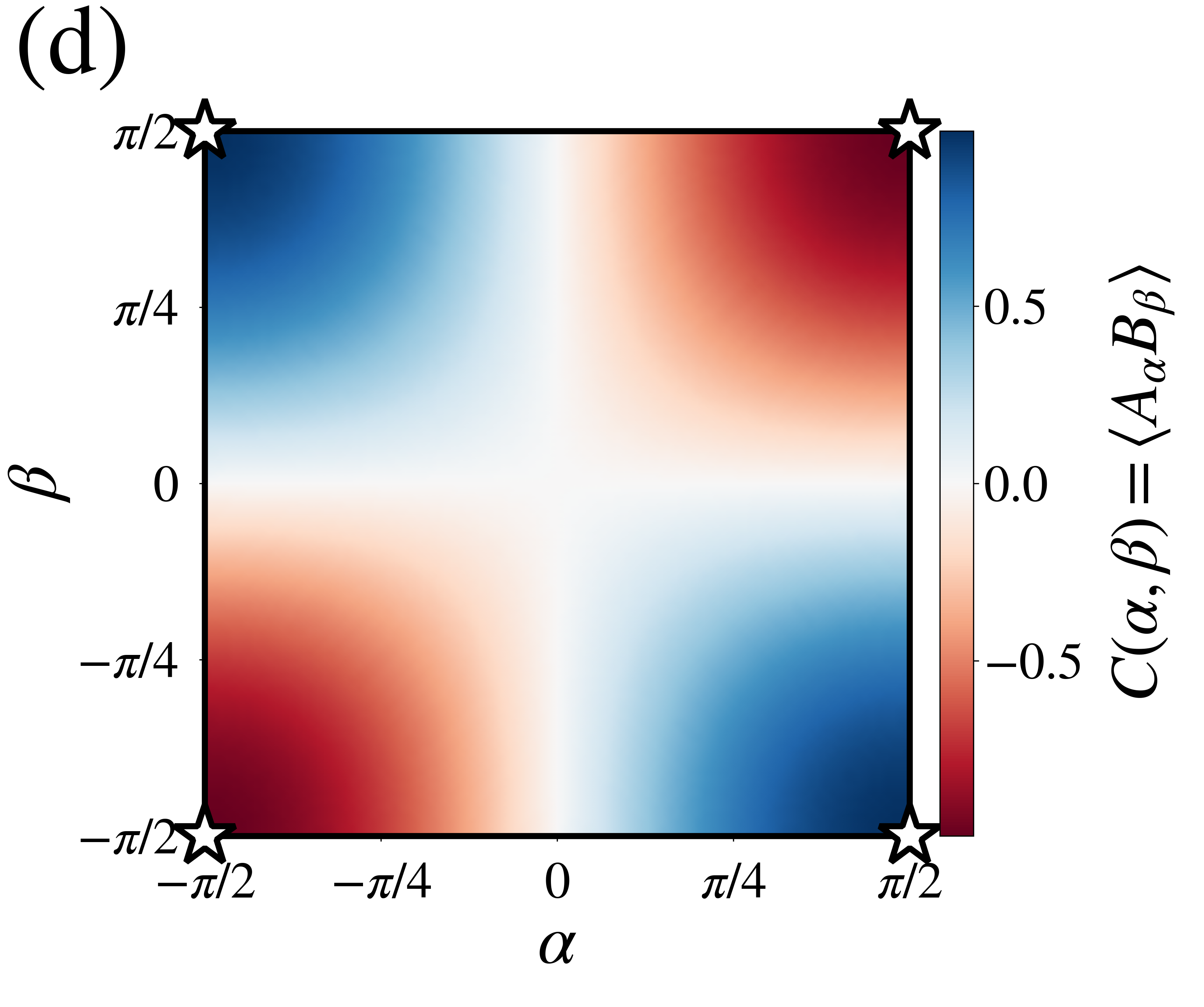}
    \caption{\textbf{Separable product-state reference.}
    The communities $(\kzz,\kzo,\koz,\koo)$ each contain $n=50$ vertices, with internal regularities $(k_1,k_2,k_3,k_4)=(21,30,30,25)$ and inter-community biregularities $(l_{12},l_{13},l_{14},l_{23},l_{24},l_{34})=(14,14,14,7,12,12)$.
    \textbf{(a)} Four-community network with positive couplings.
    \textbf{(b)} Spectrum with an isolated, nondegenerate dominant eigenvalue.
    \textbf{(c)} Equal positive vertex amplitudes identify $|\Phi_{\mathrm{prod}}\rangle=(\kzz+\kzo+\koz+\koo)/2$.
    \textbf{(d)} Correlations reconstructed from ten growth-rate shifts factorize into independent functions of $\alpha$ and $\beta$, as expected for a product state. Stars mark the settings maximizing the reconstructed CHSH value on the sampled grid, giving $|S|=1.9877\leq2$.}
    
    \label{fig:product}
\end{figure}

We apply this reconstruction to the encoded Bell state of Eq.~\eqref{bellstate} and, as a separable reference, to the product state $|\Phi_{\rm prod}\rangle=(\kzz+\kzo+\koz+\koo)/2$. Their ideal correlations are $C_{\Psi^+}(\alpha,\beta)=-\cos(\alpha-\beta)$ and $C_{\rm prod}(\alpha,\beta)=-\sin\alpha\,\sin\beta$, respectively. The reconstructed landscapes for the two networks are shown in Figs.~\ref{fig:heat} and \ref{fig:product}(d).

To evaluate the CHSH parameter for the Bell-like state, we choose $\alpha_0=0$, $\alpha_1=\pi/2$, $\beta_0=\pi/4$, and $\beta_1=-\pi/4$, and combine the four corresponding correlations as
\begin{equation}
    S=C(\alpha_0,\beta_0)+C(\alpha_0,\beta_1)+C(\alpha_1,\beta_0)-C(\alpha_1,\beta_1).
\end{equation}
The reconstructed network statistics give the CHSH functional $|S|=\chsh$, above the CHSH threshold $2$ \cite{CHSH1969} and close to the ideal Bell-state value $2\sqrt{2}$ \cite{cirelson1980quantum}. The violation serves as a signature of the reconstructed Bell-type correlations rather than a test of Bell nonlocality.

As a separable control, we apply the same reconstruction procedure to the product-state network and search over the four measurement angles $\alpha_0$, $\alpha_1$, $\beta_0$, and $\beta_1$ for the largest CHSH value. The maximum obtained over the sampled parameter space is $|S|=1.9877\leq2$, consistent with the CHSH bound for a separable state.

Therefore, the contrast between the two networks identifies the Bell-like collective mode as an emergent entanglement-like state within the effective two-qubit description of the classical network.


{\it Discussion.---}Our dynamical readout is independent of the particular network encoding used to represent the effective state. While the present construction represents two qubits through four collective communities, other architectures can realize multiqubit state spaces differently, including Cartesian-product networks \cite{scholes2025product}. Once an effective code subspace is identified, a complete set of connectivity perturbations can probe the encoded state through its spectral responses, from which the corresponding measurement statistics are reconstructed. The dynamical readout is therefore encoding-agnostic and is not restricted to the four-community realization considered here.

{\it Summary.---}We have shown that a four-community classical network can realize and probe an effective two-qubit state through its collective dynamics and connectivity. The dominant collective mode encodes the state, while ten elementary connectivity perturbations provide an informationally complete set of spectral responses for the real two-qubit sector. From these responses, joint-outcome probabilities associated with arbitrary real projectors are reconstructed by linear combination, and the same data generate correlations over a continuous family of measurement settings. As a benchmark, the encoded Bell state gives a reconstructed CHSH functional $|S|=\chsh>2$, whereas the separable product-state reference remains within the CHSH bound. Our results therefore establish a network-native dynamical readout that extends the quantum--classical correspondence from state encoding to measurement statistics.




{\it Acknowledgments.---}This work was supported by the U.S. Department of Energy (DOE), Office of Basic Energy Sciences, under Grant No.~DE-SC0026309; the U.S. Department of Energy, Office of Basic Energy Sciences, through the DE-SC0025620 EFRC award of the Quantum Photonic Integrated Design Center (QuPIDC); and the Quantum Science Center, a National Quantum Information Science Research Center of the U.S. Department of Energy (DOE), operated at Oak Ridge National Laboratory (ORNL).

\bibliography{refs}
\end{document}

%% file: macros.tex
\usepackage[dvipsnames]{xcolor}

\usepackage{thmtools}

\declaretheorem[sibling=theorem]{
    corollary,
    conjecture,
    proposition,
    lemma,
    definition,
    example
}

\DeclarePairedDelimiter\ket{\lvert}{\rangle}
\DeclarePairedDelimiterX\braket[2]{\langle}{\rangle}{#1 \delimsize\vert #2}
\DeclarePairedDelimiterX\braketnom[2]{\langle}{\rangle}{#1 #2}
\DeclarePairedDelimiterX\ketbra[2]{\delimsize\vert}{\delimsize\vert}{#1 \rangle \langle #2}

\newcommand{\kzz}{\ket{00}}
\newcommand{\kzo}{\ket{01}}
\newcommand{\koz}{\ket{10}}
\newcommand{\koo}{\ket{11}}

\newcommand{\bellp}{\Psi^+}

\newcommand{\kbellp}{\ket{\bellp}}

\newcommand{\onesqrttwo}{\frac{1}{\sqrt{2}}}

\usepackage{xspace}

